\documentclass{article}

\usepackage{graphicx} \usepackage{amsmath} \usepackage{amsfonts}
\usepackage{amssymb}

\begin{document}

\thispagestyle{empty}

\title{Four dimensional Lie symmetry algebras and fourth order
ordinary differential equations}

\label{firstpage}

\author{T Cerquetelli~$^\dag$, N Ciccoli~$^\dag$ and MC
Nucci~$^{\dag,\ddag}$}

\date{$^\dag$~ Dipartimento di Matematica e Informatica,
Universit\`a di Perugia, 06123 Perugia, Italy \\[10pt] $^\ddag$~
Corresponding author. E-mail: nucci@unipg.it}

\maketitle

\begin{abstract}
\noindent Realizations of four dimensional Lie algebras as vector
fields in the plane are explicitly constructed. Fourth order
ordinary differential equations which admit such Lie symmetry
algebras are derived. The route to their integration is described.
\end{abstract}

\section{Introduction}
In the second half of the XIX century,  Marius Sophus Lie
(1842-1899), the great Norwegian mathematician, studied a class of
special algebras, which he called continuous groups of
transformations. In the Preface to his book \cite{bianchi}, Luigi
Bianchi (1856-1928), an Italian mathematician, who was a
contemporary of Lie, wrote an eulogy about the colleague's work:
 \begin{verse} \noindent Movendo da concetti geometrici,
associati allo studio dei problemi d'integrazione, Egli riconobbe
l'importanza fondamentale, per la geometria e per l'analisi, della
considerazione di questi gruppi {\em continui}, e concep\`{\i} ed
attu\`o l'ardito disegno di co\-struir\-ne la teoria
ge\-ne\-ra\-le che doveva estendere al campo continuo la teoria
dei gruppi di sostituzioni e quivi compiere, per le teorie
d'integrazione nell'analisi, un'opera di {\em classificazione}
analoga a quella della teoria di  GALOIS nello studio delle
irrazionalit\`a algebriche.\\ \noindent E per opera di S. LIE la
teoria dei gruppi continui, per quanto riguarda i gruppi {\em
finiti} (che dipendono cio\`e da un numero finito di parametri),
venne completamente costituita, arricchendo la scienza matematica
di una delle pi\`u importanti conquiste del secolo
scorso.\footnote{He moved from geometric concepts that are
associated with the study of problems of integration in order to
recognize the fundamental importance of considering those {\em
continuous} groups in geometry and analysis, and to conceive and
realize the daring plan of constructing their general theory which
was going to extend the theory of substitution groups to the
continuous field, and thus complete for the theories of
integration in analysis a {\em classification} work  similar to
that of GALOIS' theory in the study of algebraic irrationalities.
Thanks to S LIE the theory of continuous groups, for what concerns
the {\em finite} groups (those which depend by a finite number of
parameters), was completely constituted; that  enriched the
mathematical science of one of the most important achievements of
the last century.}
\end{verse}
Upon Lie's death Bianchi wrote an obituary \cite{bianchi99} in
which he describes Lie's work. In particular, he said:
\begin{verse} \noindent Il LIE determin\`o inoltre tutti i
possibili tipi di gruppi finiti continui sopra una, due  o tre
variabili o, se si vuole, sulla retta, nel piano o nello spazio
(in quest'ultimo caso soltanto, in modo completo, pei gruppi
primitivi).\footnote{Moreover LIE determined all the possible
finite continuous groups on one, two, three variables or, as one
wishes, on the line, plane or space (in the latter case, only for
primitive groups completely)}\end{verse} Today the problem of
determining all finite dimensional continuous groups is formulated
as the problem of determining all finite dimensional Lie algebras
of vector fields up to equivalence under diffeomorphisms
\cite{Hawkins}. A recent account on the problem of classifying Lie
algebras of vector fields can be found in \cite{olver}. There
Gonz\'alez-L\'opez et al. give the classification of Lie algebras
of differential operators in two real variables. Also they state
that {\em ``Lie, Campbell, Bianchi, etc., never really made it
clear whether they were working over the real or the complex
numbers''}. Let us render justice to Bianchi for he did
distinguish between complex and real space. In fact in
\cite{bianchi97} Bianchi gave the classification of all the real
algebras of vector fields in the real space. He based his work on
Lie's classification, but stated that
\begin{verse} \noindent Nella
classificazione di LIE non vi \`e luogo a distinguere il reale
dall'immaginario, laddove noi vogliamo, in queste ricerche,
riferirci soltanto a gruppi reali ed ai loro sottogruppi reali:
dovremo perci\`o suddividere in pi\`u tipi qualche tipo, che dal
punto di vista generale del LIE risulta unico.\footnote{In LIE's
classification there is no place for distinguishing the real from
the imaginary; whereas  in the present research, we want to refer
to real groups and their real subgroups only: therefore we shall
have to subdivide into more types certain types, which are unique
from LIE's general point of view.}\end{verse}
  In particular, he
introduced the Type IX three-dimensional Lie algebra which does
not contain any two-dimensional subalgebra:
\begin{verse}
\noindent  Resta infine da considerare il caso in cui il gruppo
$G_3$ non \`e integrabile. Per questi gruppi LIE assegna l'unico
tipo\\ \noindent
 (Tipo VIII) $\;\;\;\;\;
 (X_1\,X_2)=X_1f,\;(X_1\,X_3)=2X_2f,\;(X_2\,X_3)=X_3f,$\\
 \noindent
 ma noi dovremo aggiungervi
l'altro:\\ \noindent
 (Tipo IX)
 $\;\;\;\;\;(X_1\,X_2)=X_3f,\;(X_2\,X_3)=X_1f,\;(X_3\,X_1)=X_2f,$\\
 \noindent
il quale differisce dal precedente per ci\`o che in quest'ultimo
non esiste alcun sottogruppo {\em reale} a due
parametri.\footnote{Finally it remains to consider the case in
which  group $G_3$ is not integrable. For these groups LIE gives a
unique type (Type VIII) $\ldots$, but we will have to add another
one (Type IX) $\ldots$, which differs from the other because in
the latter it does not exist any {\em real} subgroup with two
parameters.}
 \end{verse}
  In \cite{bianchi97} Bianchi also proved the following
  \begin{verse} \noindent
dimostriamo che un gruppo (transitivo) di movimenti con 6, con 5
ovvero con 4 parametri contiene necessariamente qualche
sottogruppo {\em reale} a 3 parametri.\footnote{we show that  a
(transitive) group of motions with either 6, 5 or 4 parameters
necessarily contains some {\em real} subgroups with 3 parameters.}
\end{verse}
In this paper we use this result to construct a (possible)
realization in the plane of  four-dimensional Lie algebras by
considering one of their  subalgebras which were listed by Patera
and Winternitz in \cite{PW}. In particular we take into
consideration the realizations of three-dimensional Lie algebras
in the plane derived by  Mahomed in \cite{mahomed}. Moreover we
determine fourth order ordinary differential equations admitting
those realizations as their Lie symmetry algebra.
 Finally we show the route to integration.

\section{Four-dimensional Lie algebras of vector fields}

 Let ${\mathfrak g}$ be a real Lie algebra. A realization of
 ${\mathfrak g}$ as vector fields in the plane is an injective Lie algebra morphism
 $T:{\mathfrak g}\to{\mathfrak X}({\mathbb R}^2)$.
It is common to identify the realization $T$ with its image. Let
us remark that for any given realization of a Lie algebra one has
a naturally defined effective local Lie action on ${\mathbb R}^2$.

To achieve the general goal of realizing finite dimensional Lie
algebras one can, in principle, proceed by induction on the
dimension of ${\mathfrak g}$. Having classified  realizations up
to dimension $(n-1)$ one can consider in every $n$--dimensional
Lie algebra a maximal Lie subalgebra together with one of its
realizations (there can be more than one in general) and try to
extend it to a realization of the whole algebra. This means
imposing the commutators on a generic vector field which
translates into a system of linear partial differential equations.
The explicit solutions of such system, if any, provide
realizations of the algebra. However, such a programme encounters
two major difficulties: neither the classification of
$n$--dimensional Lie algebras nor the classification of maximal
subalgebras of a given Lie algebra are known.

In what follows we restrict to the case of four-dimensional Lie
algebras, a situation in which both classifications are well known
and explicit. We rely upon \cite{PW} from which we have  borrowed
notations as well: $A_{p,q}$ denotes a Lie algebra of dimension
$p$ and isomorphism type $q$. The classification results are
recollected in Table 1. Realizations of three-dimensional Lie
algebras, which are our building brick, can be found in
\cite{mahomed}.

 If we try to find a realization of a four-dimensional
algebra, then we need to consider a realization  of one of its
three-dimensional subalgebras. Thus we have to determine only the
remaining operator
\begin{equation}\label{opg}
e_{j}=a_{j}(x,y)\frac{\partial}{\partial
x}+b_{j}(x,y)\frac{\partial}{\partial
y}\;\;\;\;\;\;\;(j=1,\ldots,q \leq 4)
\end{equation}
with $a_{j}$ and $b_{j}$ arbitrary functions of $(x,y)$. Imposing
the commutation relations which characterize the four-dimensional
algebra generates an overdetermined  linear system of partial
differential equations in the two unknowns $a_j$ and $b_j$. The
solution of this system may lead to a realization of the algebra.
We give an example.
\begin{center} {\bf Algebra} $ A_{4,12}$ \end{center}
Consider the three-dimensional subalgebra, $ \ A_{3,3}^{II}$:($
e_{3};e_{1},e_{2}$), which has the following realization
\cite{mahomed}
\begin{equation}\label{4121}
e_{1} =  \displaystyle{\frac{\partial}{\partial y}},\;\;\; \;\;
e_{2}
 =  x\displaystyle{\frac{\partial}{\partial y}},\;\;\;\; \; e_{3} =
y\displaystyle{\frac{\partial}{\partial y}}
\end{equation}
Let $ \ e_{4}$ be an operator of type (\ref{opg}). If we rerquire
that the operators   (\ref{4121}) and (\ref{opg}) satisfy the
commutation relations of $ \ A_{4,12}$ (see Table 1), then we
obtain
\begin{equation}\label{412}
e_{1} =  \displaystyle{\frac{\partial}{\partial y}},\;\; \;\;\;
e_{2}
 = x\displaystyle{\frac{\partial}{\partial y}},\;\;\;\; \; e_{3} =
y\displaystyle{\frac{\partial}{\partial y}},\;\;\; \;\; e_{4} =
-(1+x^{2})\displaystyle{\frac{\partial}{\partial
x}}-xy\displaystyle{\frac{\partial}{\partial y}}.
\end{equation}

This procedure can be repeated in each case and all the results
are listed in Table 2. Due to the complexity of such systems the
computations were carried out using  REDUCE 3.7, a computer
algebra software. However, the algebras of type $A_2\oplus 2A_1,
A_{3,1}\oplus A_1, A_{3,3}\oplus A_1, A_{3,9}\oplus A_1,
A_{4,2}^1, A_{4,5}^{a,a}, A_{4,5}^{a,1}, A_{4,10}, A_{4,11}^a$
yield incompatible systems.  Thus we infer that they cannot be
realized in the plane.

\section{Fourth order equations admitting a four-dimensional Lie symmetry algebra}

Having a realization of a four-dimensional real Lie algebra we can
construct a fourth order ordinary differential equation (ODE)
which admits such an algebra as its Lie symmetry algebra by
finding the differential invariants of the Lie algebra up to
fourth order.  We consider the most general form of a fourth order
ODE
\begin{equation}\label{eg4}
\Phi(x,y,y',y'',y''',y^{iv})=0.
 \end{equation}
We prolong the operators as given in Table 2 up to the fourth
order. Then the solution of the system
\begin{equation}\label{sp4}
e_{i}^{4}(\Phi) \mid _{\Phi =0}=0\;\;\;\;\;\;\;(i=1,2,3,4)
\end{equation}
yields the differential invariants and, obviously, the
corresponding fourth order ODE. We use {\it ad hoc} interactive
REDUCE programs developed by  one of the authors \cite{Reduce} to
perform this lengthy
 task.
All the equations we have found are listed in Table 3.  We show a
detailed example.
\begin{center} {\bf Algebra} $A_{4,12}$ \end{center}
Consider the realization (\ref{412}). The prolongations of those
operators up to fourth order yield
\begin{equation}\label{pa412}
\begin{array}{rcl}
e_{1}^{4} &= & \displaystyle{\frac{\partial}{\partial y}}\\ \\
e_{2}^{4} &= & x\displaystyle{\frac{\partial}{\partial
y}}+\displaystyle{\frac{\partial}{\partial y'}}\\ \\ e_{3}^{4} &=
& y\displaystyle{\frac{\partial}{\partial
y}}+y'\displaystyle{\frac{\partial}{\partial
y'}}+y''\displaystyle{\frac{\partial}{\partial
y''}}+y'''\displaystyle{\frac{\partial}{\partial
y'''}}+y^{iv}\displaystyle{\frac{\partial}{\partial y^{iv}}}\\ \\
e_{4}^{4} &= & -(1+x^{2})\displaystyle{\frac{\partial}{\partial
x}}-xy\displaystyle{\frac{\partial}{\partial y
}}+(xy'-y)\displaystyle{\frac{\partial}{\partial
y'}}+3xy''\displaystyle{\frac{\partial}{\partial
y''}}+(3y''+5xy''')\displaystyle{\frac{\partial}{\partial
y'''}}+\\
\\ &\;&
+(7xy^{iv}+8y''')\displaystyle{\frac{\partial}{\partial y^{iv}}}.
\end{array}
\end{equation}
Now we solve the corresponding system (\ref{sp4}).  The first and
second equations imply that  (\ref{eg4}) does not depend on
$y,y'$, i.e.
\begin{equation}\label{eg412}
\Phi(x,y'',y''',y^{iv})=0.
\end{equation}
In order to integrate the third equation, we must solve the
following equations for the characteristics:
\begin{equation}\label{c412}
\frac{{\rm d}y''}{y''}=\frac{{\rm d}y'''}{y'''}=\frac{{\rm
d}y^{iv}}{y^{iv}}.
\end{equation}
We obtain the following differential invariants:
\begin{equation}\label{i412}
I_{1}=\frac{y'''}{y''},\;\;\;I_{2}=\frac{y^{iv}}{y''}
\end{equation}
which force equation  (\ref{eg412}) to become
\begin{equation}\label{eg412p}
\Phi(x,I_{1},I_{2})=0.
\end{equation}
Finally, after the substitution of the invariants (\ref{i412}),
the fourth equation  is integrated by solving the following
characteristic equations:
\begin{equation}\label{c4122}
\frac{{\rm d}x}{-(1+x^{2})}=\frac{{\rm
d}I_{1}}{3+2xI_{1}}=\frac{{\rm d}I_{2}}{4xI_{2}+8I_{1}}
\end{equation}
which yield the following differential invariants:
\begin{equation}\label{x412}
J_{1}=\frac{y'''(1+x^{2})+3xy''}{y''},\;\;\;\;\;
J_{2}=\frac{(1+x^{2})[(1+x^{2})y^{iv}+8xy''']+12{x}^{2}y''}{y''}.
\end{equation}
Then equation (\ref{eg412p}) becomes
\begin{equation}\label{eg412s}
\Phi(J_{1},J_{2})=0,
\end{equation}
videlicet, by Dini's theorem:
\begin{equation}\label{ea412}
y^{iv}=\frac{\displaystyle y''F(J_1)-8xy'''(1+x^2)-12x^{2}y''}
{\displaystyle (1+x^{2})^{2}},
\end{equation}
with $F$ an arbitrary function of $J_1$. Equation (\ref{ea412})
admits $ \ A_{4,12} $ as its Lie symmetry algebra.

\section{Integration of fourth order equations by Lie's method}

Lie showed that  an ordinary differential equation of order $n$
with a known $n$--dimensional Lie symmetry algebra can be
integrated by quadratures provided that its symmetry algebra is
solvable \cite{Lie5}. The general integrating procedure consists
of $n$ successive integrations and leads to quite lengthy
calculations. Among the equations in Table 3 only $A_{3,8}\oplus
A_1$ is not solvable. In \cite{nailio} the integrating procedure
was provided for any third order ODE which admits either a
solvable or not solvable three-dimensional Lie symmetry algebra
${\mathfrak g}_{3}$. If ${\mathfrak g}_{3}$ is solvable, then we
can reduce the given third order equation to a first order ODE
which is integrable by quadrature and then obtain a second order
ODE which can be transformed into a directly integrable form
(Lie's method). If ${\mathfrak g}_{3}$  is not solvable, then we
can still reduce the given third order equation to a first order
equation; this equation is not integrable by quadrature,  but can
be easily reduced to a Riccati equation\footnote{In
\cite{clarkson} a theoretical explanation of the appearance of a
Riccati equation was given.} by using a nonlocal symmetry which
comes from one of the symmetries of the original third order ODE.

 Here we follow a similar procedure.  Consider a fourth order ODE which admits a four-dimensional
solvable Lie algebra ${\mathfrak g}_{4}$. Firstly we reduce it to
a first order ODE by using the differential invariants of an ideal
${\mathfrak h}_{3} \subset {\mathfrak g}_{4}$. Then the first
order equation can be integrated by quadrature because it admits
the one-dimensional Lie algebra ${\mathfrak g}_{4}/{\mathfrak
h}_{3}$. Its general solution becomes a third order ODE in the
original variables. This equation admits ${\mathfrak h}_{3}$.
Therefore it can be integrated with the procedure showed in
\cite{nailio}. If a fourth-order ODE admits a Lie symmetry algebra
${\mathfrak g}_{4}$ which is not solvable, then we can always
reduce it to a first order ODE by using the differential
invariants of a three-dimensional subalgebra ${\mathfrak g}_{3}$.
Finally the first order ODE can be integrated by using a nonlocal
symmetry which comes from the fourth symmetry.  We show in details
the case of a fourth order equation which admits a solvable Lie
symmetry algebra and that of the equation which admits
$A_{3,8}\oplus A_1$ as its Lie symmetry algebra.

\begin{center} {\bf Algebra} $ A_{4,12}$ \end{center} Consider the
realization
 \begin{equation}\label{inta412}
e_{1}=\frac{\partial}{\partial
y},\;\;\;\;e_{2}=x\frac{\partial}{\partial
y},\;\;\;\;e_{3}=y\frac{\partial}{\partial
y},\;\;\;\;e_{4}=-(1+x^{2})\frac{\partial}{\partial
x}-xy\frac{\partial}{\partial y}
 \end{equation}
and the fourth order ODE  which admits $A_{4,12}$ with generators
(\ref{inta412}) as its Lie symmetry algebra
 \begin{equation}\label{a412e}
y^{iv}=\frac{\displaystyle y''F(\xi)-8xy'''(1+x^2)-12x^{2}y''}
{\displaystyle (1+x^{2})^{2}},\;\;\; \xi=\frac{\displaystyle
y'''(1+x^{2})+3xy''}{\displaystyle y''}.
 \end{equation}
 The commutation relations are: $$
[e_{1},e_{3}]=e_{1},\;\;\;[e_{2},e_{3}]=e_{2},\;\;\;[e_{1},e_{4}]=-e_{2},\;\;\;
[e_{2},e_{4}]=e_{1}.
$$
The algebra is solvable and the operators $e_{1},e_{2},e_{3}$
generate a three-dimensional ideal ${\mathfrak h}_{3}=\langle
e_{1},e_{2},e_{3}\rangle$. A basis of its differential invariants
of order $\leq 3$ is:
 \begin{equation}\label{inva412}
u=x,\;\;\;\;\;\;\;\;v=\frac{y'''}{y''}
 \end{equation}
Then equation (\ref{a412e}) can be reduced to the first order ODE
\begin{equation}\label{a412uv}
\frac{{\rm d}v}{{\rm
d}u}=\frac{F(\tilde\xi)-8uv(1+u^{2})-12u^{2}-v^{2}(1+u^{2})^{2}}{(1+u^{2})^{2}},\;\;
\;\;\;\;\tilde\xi=3u+v(1+u^{2})
 \end{equation}
which admits the one-dimensional Lie symmetry algebra generated by
 $$
e_{4}=-(1+u^{2})\frac{\partial}{\partial
u}+(3+2uv)\frac{\partial}{\partial v}\, .
 $$
We write equation (\ref{a412uv}) as a linear differential form
 \begin{equation}\label{fdl412}
(F(\tilde\xi)-8uv(1+u^{2})-12u^{2}-v^{2}(1+u^{2})^{2}){\rm
d}u-(1+u^{2})^{2}{\rm d}v=0.
 \end{equation}
Its integrating factor is \cite{Lie5}
 $$
I=-\frac{1}{(1+u^{2})(F(\tilde\xi)-9u^{2}+3-v(1+u^{2})(6u+1+u^{2}))}.
 $$
Therefore the general integral of equation (\ref{fdl412}) is
obtained in the form
 $U(u,v)=c_1$  by the solution of
 \begin{equation}\label{fa412}
\left\{ \begin{array}{rcl} \displaystyle{\frac{\partial
U}{\partial u}} &= &
-\displaystyle{\frac{F(\tilde\xi)-8uv(1+u^{2})-12u^{2}-v^{2}(1+u^{2})^{2}}{(1+u^{2})
(F(\tilde\xi)-9u^{2}+3-v(1+u^{2})(6u+1+u^{2}))}};\\
\\ \displaystyle{\frac{\partial U}{\partial v}} &= &
-\displaystyle{\frac{(1+u^{2})}{F(\tilde\xi)-9u^{2}+3-v(1+u^{2})(6u+1+u^{2})}};\\
\end{array} \right.\, .
 \end{equation}
Substitution of the original variables into $U$ yields a third
order ODE of the form $ U\left(x,\frac{y'''}{y''}\right)=c_1$
which admits the Lie symmetry algebra generated by
$e_{1}={\partial_ y}$, $e_{2}=x{\partial_ y}$,
$e_{3}=y{\partial_y}$ and can be solved by quadrature
\cite{nailio}.

\begin{center} {\bf Algebra} $ A_{3,8}\oplus A_1$ \end{center}
Consider the realization
 \begin{equation}\label{inta38a1}
e_{1}=\frac{\partial}{\partial
y},\;\;\;e_{2}=y\frac{\partial}{\partial
y},\;\;\;e_{3}=-y^2\frac{\partial}{\partial
y},\;\;\;e_{4}=\frac{\partial}{\partial x}
 \end{equation}
and the fourth order equation which admits $ A_{3,8}\oplus A_1$
with the generators (\ref{inta38a1}) as its Lie symmetry
algebra\footnote{Of course it is not a surprise that $\xi$
corresponds to the Schwarzian derivative \cite{nailio}.}
 \begin{equation}\label{a38a1e}
y^{iv}=\frac{\displaystyle-3{y''}^{3}+4y'y''y'''
  +{y'}^3F(\xi)}{\displaystyle {y'}^2},\;\;\;\;\;
  \xi =\frac{\displaystyle y'''}
  {\displaystyle y'}-\frac{\displaystyle 3{y''}^{2}}{\displaystyle 2{y'}^{2}}.
 \end{equation}
 The commutation relations are: $$
[e_1,e_3]=-2e_2,\;\;\;\;\;\;[e_2,e_3]=e_3,\;\;\;\;\;\;[e_1,e_2]=e_1.
$$
The algebra is not solvable, but the operators $e_{1},e_{2},e_{3}$
generate a three-dimensional subalgebra ${\mathfrak g}_{3}=\langle
e_{1},e_{2},e_{3}\rangle$. A basis of its differential invariants
of order $\leq 3$ is:
 \begin{equation}\label{inva38a1}
u=x,\;\;\;\;\;\;\;\;v=\frac{y'''}{y'}-\frac{3}{2}\frac{{y''}^2}{{y'}^2}.
 \end{equation}
Then equation (\ref{a38a1e}) can be reduced to the first order ODE
\begin{equation}\label{a38a1uv}
\frac{{\rm d}v}{{\rm d}u}=F(v)
 \end{equation}
which admits the one-dimensional Lie symmetry algebra generated by
 $$
e_{4}=\frac{\partial}{\partial u}.
 $$
Therefore (\ref{a38a1uv}) can be easily integrated by quadrature,
i.e.
$$\int \frac{{\rm d} v}{F(v)} =u+c_1$$
which in the original variables becomes a third order ODE which
admits ${\mathfrak g}_{3}$ as its Lie symmetry algebra and can
then be integrated \cite{nailio}.

\section{Tables}
In Table 1 we list the four-dimensional real Lie algebras as given
in \cite{PW}. In the second column the nonzero commutation
relations are given.  In the last column the suitable three
dimensional subalgebra that we have used either to generate a
realization or  to disprove that a realization exists -- algebras
marked with $(\ast)$ -- are specified.\\

\noindent  In Table 2 we have put the realizations that we found,
with $f,f_1,f_2$ arbitrary functions. The generators of each
algebra are orderly listed as $e_1, e_2,
e_3, e_4$.\\

\noindent In Table 3 we list the fourth order ODEs which admit one
of Lie algebras in Table 2 as Lie symmetry algebras. Note that $F$
represents an arbitrary function and that we have chosen a
particular form -- shown in the third column -- for each of the
arbitrary functions listed in Table 2.\\

\newpage
\begin{center}
{\small Table 1\\}
 \begin{tabular}{|l|l|l|}
  \hline \hline
  Lie algebra & Nonzero commutation relations & Subalgebra\\ \hline \hline
  $4A_1$ & & $3A_1:\langle e_2,e_3,e_4  \rangle$ \\
  \hline
 $(\ast)$ $A_2\oplus 2A_1$ & $[e_1,e_2]=e_2$ & $3A_1:\langle e_2,e_3,e_4\rangle$  \\
   \hline
  $2A_2$ & $[e_1,e_2]=e_2,[e_3,e_4]=e_4$ & $A_1\oplus A_2:\langle e_1,e_4,e_2\rangle$ \\
   \hline
 $(\ast)$  $A_{3,1}\oplus A_1$ & $[e_2,e_3]=e_1$ & $3A_1:\langle e_1,e_2,e_4\rangle$ \\
  \hline
$A_{3,2}\oplus A_1$ & $[e_1,e_3]=e_1,[e_2,e_3]=e_1+e_2$ & $3A_1:\langle e_1,e_2,e_4\rangle $ \\
\hline
  $(\ast)$ $A_{3,3}\oplus A_1$  & $[e_1,e_3]=e_1,[e_2,e_3]=e_2$ & $3A_1:\langle e_1,e_2,e_4\rangle$ \\
  \hline
  $A_{3,4}\oplus A_1$ & $[e_1,e_3]=e_1,[e_2,e_3]=-e_2$ & $3A_1:\langle e_1,e_2,e_4\rangle$ \\
  \hline
  $A_{3,5}^a\oplus A_1$ & $[e_1,e_3]=e_1,[e_2,e_3]=ae_2$ & $3A_1:\langle e_1,e_2,e_4\rangle$
  \\
   $(0<|a|<1)$& & \\ \hline
  $A_{3,6}\oplus A_1$ & $[e_1,e_3]=-e_2,[e_2,e_3]=e_1$ & $3A_1:\langle e_1,e_2,e_4\rangle$ \\
  \hline
$A_{3,7}^a\oplus A_1$ $(a>0)$
&$[e_1,e_3]=ae_1-e_2,[e_2,e_3]=e_1+ae_2$ & $3A_1:\langle e_1,e_2,e_4\rangle$ \\
 \hline
 $A_{3,8}\oplus A_1$ &
$[e_1,e_3]=-2e_2,[e_2,e_3]=e_3,[e_1,e_2]=e_1$ & $A_{3,8}:\langle
e_1,e_2,e_3\rangle$ \\\hline $(\ast)$   $A_{3,9}\oplus A_1$  &
$[e_1,e_3]=-e_2,[e_2,e_3]=e_1,[e_1,e_2]=e_3$
  & $A_{3,9}:\langle e_1,e_2,e_3\rangle$ \\\hline
  $A_{4,1}$ & $[e_2,e_4]=e_1,[e_3,e_4]=e_2$ & $3A_1:\langle e_1,e_2,e_3\rangle$  \\\hline
  $A_{4,2}^a$ $(a\neq 0,1)$& $[e_1,e_4]=ae_1,[e_2,e_4]=e_2,[e_3,e_4]=e_2+e_3$ &
    $3A_1: \langle e_1,e_2,e_3\rangle$  \\ \hline
$(\ast)$   $A_{4,2}^1$  &
$[e_1,e_4]=e_1,[e_2,e_4]=e_2,[e_3,e_4]=e_2+e_3$  &
   $3A_1: \langle e_1,e_2,e_3\rangle$  \\ \hline
  $A_{4,3}$ & $[e_1,e_4]=e_1,[e_3,e_4]=e_2$ & $3A_1:\langle e_1,e_2,e_3\rangle$   \\\hline
  $A_{4,4}$ & $[e_1,e_4]=e_1,[e_2,e_4]=e_1+e_2,[e_3,e_4]=e_2+e_3$ &
  $3A_1:\langle e_1,e_2,e_3\rangle$   \\ \hline
  $A_{4,5}^{a,b}$ $(ab\neq 0)$& $[e_1,e_4]=e_1,[e_2,e_4]=ae_2,[e_3,e_4]=be_3$ &
   $3A_1:\langle e_1,e_2,e_3\rangle$   \\
   $(-1\leq a<b<1$  & &\\\hline
 $(\ast)$  $A_{4,5}^{a,a}$ & $[e_1,e_4]=e_1,[e_2,e_4]=ae_2,[e_3,e_4]=ae_3$ &
  $3A_1:\langle e_1,e_2,e_3\rangle$   \\
  $(-1\leq a<1,a\neq 0)$ & &\\\hline
$(\ast)$   $A_{4,5}^{a,1}$ &
$[e_1,e_4]=e_1,[e_2,e_4]=ae_2,[e_3,e_4]=e_3$
  & $3A_1:\langle e_1,e_2,e_3\rangle$   \\
  $(-1\leq a<1,a\neq 0)$ & &\\\hline
  $A_{4,5}^{1,1}$ & $[e_1,e_4]=e_1,[e_2,e_4]=e_2,[e_3,e_4]=e_3$ &
   $3A_1:\langle e_1,e_2,e_3\rangle$   \\\hline
  $A_{4,6}^{a,b}$ $(a\neq 0, b\geq 0)$ & $[e_1,e_4]=ae_1,[e_2,e_4]=be_2-e_3,[e_3,e_4]=e_2+be_3$
  & $3A_1:\langle e_1,e_2,e_3\rangle$   \\
  \hline
  $A_{4,7}$ & $[e_1,e_4]=2e_1,[e_2,e_4]=e_2,$
     & $A_{3,5}^{1/2}:\langle e_1,e_2,e_4\rangle$   \\&$[e_3,e_4]=e_2+e_3,[e_2,e_3]=e_1$&\\\hline
  $A_{4,8}$ & $[e_2,e_3]=e_1,[e_2,e_4]=e_2,[e_3,e_4]=-e_3$
   & $A_{3,1}:\langle e_1,e_2,e_3\rangle$   \\\hline
  $A_{4,9}^b$ $(0<|b|<1)$& $[e_1,e_4]=(1+b)e_1,[e_2,e_4]=e_2,$
    & $A_{3,1}:\langle e_1,e_2,e_3\rangle$   \\&$[e_3,e_4]=be_3,[e_2,e_3]=e_1$&\\
   \hline
  $A_{4,9}^1$ & $[e_1,e_4]=2e_1,[e_2,e_4]=e_2,$
     & $A_{3,1}:\langle e_1,e_2,e_3\rangle$   \\&$[e_3,e_4]=e_3,[e_2,e_3]=e_1$&\\\hline
  $A_{4,9}^0$ & $[e_2,e_3]=e_1,[e_1,e_4]=e_1,[e_2,e_4]=e_2$
  & $A_{3,1}:\langle e_1,e_2,e_3\rangle$   \\\hline
 $(\ast)$  $A_{4,10}$  & $[e_2,e_3]=e_1,[e_2,e_4]=-e_3,[e_3,e_4]=e_2$
  & $A_{3,1}:\langle e_1,e_2,e_3\rangle$   \\\hline
  $(\ast)$  $A_{4,11}^a$  $(0<a)$&
  $[e_1,e_4]=2ae_1,[e_2,e_4]=ae_2-e_3,$
      & $A_{3,1}:\langle e_1,e_2,e_3\rangle$   \\ & $[e_3,e_4]=e_2+ae_3,[e_2,e_3]=e_1$&\\\hline
 $A_{4,12}$ & $[e_1,e_4]=-e_2,[e_2,e_4]=e_1,[e_1,e_3]=e_1,$ &
 $A_{3,3}:\langle e_1,e_2,e_3\rangle$   \\ &$[e_2,e_3]=e_2$ &\\\hline\hline
\end{tabular}
\end{center}

\newpage

\begin{center}
{\small Table 2\\}
\begin{tabular}{|l|l|}
    \hline\hline Lie algebra & Generators \\ \hline\hline
  $4A_1$ & $f_1(x)\partial_y,\partial_y,x\partial_y,f_2(x)\partial_y$  \\
  \hline
  $2A_2$ & $-x\partial_x,\partial_x, -y\partial_y,\partial_y$  \\\hline
  $A_{3,2}\oplus A_1$ & $\partial_y,-x\partial_y, \partial_x+(y+f(x))\partial_y,e^x\partial_y$
   \\\hline
  $A_{3,4}\oplus A_1$ & $\partial_y,x^2\partial_y, x\partial_x+(y+f(x))\partial_y,x\partial_y$
   \\\hline
  $A_{3,5}^a\oplus A_1$ $(0<|a|<1)$& $\partial_y,x^{1-a}\partial_y, x\partial_x+(y+f(x))\partial_y,
  x\partial_y$\\\hline
  $A_{3,6}\oplus A_1$ & $\partial_y,(x^2-1)^{1/2}\partial_y, -x(x^2-1)^{1/2}\partial_x
  +(f(x)-y(x^2-1)^{1/2})\partial_y,  x\partial_y$\\\hline
$A_{3,7}^a\oplus A_1$ $(a>0)$& $\partial_y,x\partial_y,
-(1+x^2)\partial_x+((a-x)y+f(x))\partial_y,(1+x^2)^{1/2}
  e^{a \arctan (x)}\partial_y$\\\hline
$A_{3,8}\oplus A_1$ & $\partial_y,y\partial_y,
-y^2\partial_y,f(x)\partial_x$\\\hline $A_{4,1}$ &
$\partial_y,x\partial_y,(x^2/2)\partial_y,-\partial_x+f(x)\partial_y$\\\hline
  $A_{4,2}^a$ $(a\neq 0,1)$& $e^{(1-a)x}\partial_y,-\partial_y,x\partial_y,
  \partial_x+y\partial_y$\\\hline
  $A_{4,3}$ & $\partial_y,x\partial_y,-x\log (x)\partial_y,x\partial_x+(y+f(x))\partial_y$\\\hline
$A_{4,4}$ & $\partial_y,x\partial_y,(x^2/2)\partial_y,
-\partial_x+(y+f(x))\partial_y$\\\hline
  $A_{4,5}^{a,b}$ $(-1\leq a<b<1,ab\neq 0)$ & $\partial_y,x^{1-a}\partial_y,x^{1-b}\partial_y,
  x\partial_x+(x+y)\partial_y$\\\hline
  $A_{4,5}^{1,1}$ & $f(x)\partial_y,\partial_y,x\partial_y,y\partial_y$\\\hline
  $A_{4,6}^{a,b}$ $(a\neq 0, b\geq 0)$ & $(1+x^2)^{1/2}e^{(b-a)\arctan (x)}\partial_y,
  x\partial_y,\partial_y, (1+x^2)\partial_x+(xy+by)\partial_y$\\\hline
  $A_{4,7}$ & $\partial_y,x\partial_y,-\partial_x-x\log (x)\partial_y,
  x\partial_x+2y\partial_y$\\\hline
  $A_{4,8}$ & $\partial_y,\partial_x,x\partial_y,x\partial_x$\\\hline
  $A_{4,9}^b$ $(0<|b|<1)$ & $\partial_y,\partial_x,x\partial_y,x\partial_x+(1+b)y\partial_y$\\\hline
$A_{4,9}^1$ &
$\partial_y,\partial_x,x\partial_y,x\partial_x+2y\partial_y$\\\hline
$A_{4,9}^0$  &
$\partial_y,\partial_x,x\partial_y,x\partial_x+y\partial_y$\\\hline
  $A_{4,12}$ & $\partial_y,x\partial_y,y\partial_y,-(1+x^2)\partial_x-xy\partial_y$\\ \hline \hline
\end{tabular}
\end{center}

\newpage

\begin{center}
{\small Table 3}
\begin{tabular}{|l|l|l|}
    \hline\hline
  Lie algebra & Equation &  Functions\\ \hline\hline
$4A_1$ & $y^{iv}=F(x)$ & $f_1(x)=x^2$\\ &&$f_2(x)=x^3$\\\hline
$2A_2$ & $y^{iv}=\frac{\displaystyle {y''}^{3}}{\displaystyle
{y'}^{2}}F\left(\frac{\displaystyle y'y'''}{\displaystyle
{y''}^{2}}\right)$&\\\hline $A_{3,2}\oplus A_1$ &
$y^{iv}=(y'''-y'')F\left((y'''-y'')e^{-x}\right)+y''$
&$f(x)=0$\\\hline
 $A_{3,4}\oplus A_1$ &
$y^{iv}=x^{-3}F(x^{2}y''')$&$f(x)=0$\\\hline $A_{3,5}^a\oplus A_1$
&
$y^{iv}=\frac{\displaystyle -(a+2)x^2 y'''+F(\xi)}{\displaystyle x^{3}},$&$f(x)=0$\\
$(0<|a|<1)$ & $\xi=x^2y'''+x(a+1)y''$&\\\hline
 $A_{3,6}\oplus A_1$  & $y^{iv}=-
\frac{\displaystyle(8 x^4 y''' + 12 x^3 y'' - 10 x^2 y''' - 9 x
y'' + 2 y''') x^2 - F(\xi) }{\displaystyle(x^2 - 1)^2
 x^3},$ &$f(x)=0$\\ & $\xi=(x^2 - 1)^{1/2} (x^2 y''' + 3 x y'' - y''') x^2$  &\\\hline
 $A_{3,7}^a\oplus A_1$
& $y^{iv}=-\frac{ \displaystyle 8 x^3 y''' + 12 x^2 y'' + 8 x y'''
+ 3 y'' - a^2 y''}{\displaystyle (x^2 + 1)^2}+$ &$f(x)=0$\\
$(a>0)$ &
$\;\;\;\;\;\;\;\;\;\;+e^{-\arctan(x) a} (x^2 + 1)^{-7/2}F(\xi) ,$ & \\
& $\xi= e^{\arctan(x)
a}(x^2 + 1)^{3/2}(x^2 y''' + 3 x y'' + y''' + a y'')$ &  \\
 \hline
  $A_{3,8}\oplus A_1$ & $y^{iv}=\frac{\displaystyle-3{y''}^{3}+4y'y''y'''
  +{y'}^3F(\xi)}{\displaystyle {y'}^2},\;\;\;\;\;
  \xi =\frac{\displaystyle y'''}{\displaystyle y'}-\frac{\displaystyle 3{y''}^{2}}{\displaystyle 2{y'}^{2}}$  & $f(x)=1$\\
  \hline
  $A_{4,1}$ & $y^{iv}=F(y'''+6x)$ & $f(x)=x^3$ \\ \hline
$A_{4,2}^a$ $(a\neq 0,1)$
&$y^{iv}=(a-1)^2y''+e^xF(\xi),\;\;\;\;\;\; \xi=
\frac{\displaystyle (a-1)y''+y'''}{\displaystyle e^{x}} $
   & \\ \hline
  $A_{4,3}$ & $y^{iv}=\frac{\displaystyle 2xy''+F(xy''+x^2y''')}{\displaystyle x^{3}}$ &
  $f(x)=0$\\ \hline
  $A_{4,4}$ & $y^{iv}=y'''F(y'''e^{x})$ & $f(x)=0$\\ \hline
  $A_{4,5}^{a,b}\;\;(ab\neq 0)$ & $y^{iv}=\frac{\displaystyle -(2+a)x^{2}y'''-(1+b)(x^{2}y'''+(a+1)xy'')
  +F(\xi)}{\displaystyle x^{3}},$
  &\\ $(-1\leq a<b<1)$ &
  $\xi=x^{2}y'''+(1+a)xy''+b(xy''+ay'-a\log(x))$ &\\ \hline
$A_{4,5}^{1,1}$ & $y^{iv}=y'''F(x)$ & $f(x)=x^2$\\ \hline
$A_{4,6}^{a,b}$ & $y^{iv}=\frac{\displaystyle (a - b)^2 y'' -
(8 x^3 y''' + 12 x^2 y'' + 8 x y''' + 3 y'')}{\displaystyle (x^2 + 1)^2}+$ &\\
$(a\neq 0, b\geq 0)$ & $\;\;\;\;\;\;\;\;\;\;+ e^{\arctan(x)b}(x^2 + 1)^{-7/2} F(\xi), $ & \\
 & $\xi=(x^2 + 1)^{3/2}((a - b)y'' + x^2y''' + 3xy'' + y''')e^{-\arctan(x)b}$& \\ \hline
$A_{4,7}$ & $y^{iv}=\frac{\displaystyle e^{2y''}-1}{\displaystyle
x^{2}}F\left((xy'''-1)e^{-y''}\right)$ &\\ \hline $A_{4,8}$ &
$y^{iv}={y''}^{2}F\left(\frac{\displaystyle{y'''}^{2}}{\displaystyle{y''}^{3}}\right)$
  &\\\hline
$A_{4,9}^b$ $(0<|b|<1)$ & $y^{iv}={y''}^{(\frac{b-3}{b-1})}
F\left({y''}^{(2-b)}{y'''}^{(b-1)}\right)$ & \\ \hline
  $A_{4,9}^1$ & $ y^{iv}={y'''}^2F(y'')$ &\\ \hline
$A_{4,9}^0$ & $y^{iv}={y''}^{3}F\left(\frac{\displaystyle
y'''}{\displaystyle {y''}^{2}}\right)$ &\\\hline
 $A_{4,12}$ &
$y^{iv}=\frac{\displaystyle y''F(\xi)-8xy'''(1+x^2)-12x^{2}y''}
{\displaystyle (1+x^{2})^{2}},\;\;\; \xi=\frac{\displaystyle
y'''(1+x^{2})+3xy''}{\displaystyle y''}$&
\\\hline\hline
\end{tabular}
\end{center}

\label{lastpage}


\newpage
\begin{thebibliography}{99}
 \bibitem{bianchi97} Bianchi L, Sugli spazi a tre dimensioni che ammettono un gruppo continuo di
movimenti, {\em Mem. Soc. it. delle Sc. dei XL} (3) {\bf 11}
(1897), 267--352.
 \bibitem{bianchi99} Bianchi L, Notizie sull'opera matematica di Sophus Lie,
 {\em Rendiconti della R. Accademia dei Lincei} (5) {\bf 8} (1899), 360--366.
 \bibitem{bianchi} Bianchi L,  Lezioni sulla Teoria dei
 Gruppi Continui Finiti  di Trasformazioni, Enrico Spoerri Editore, Pisa, 1918.
 \bibitem{clarkson} Clarkson PA and Olver PJ, Symmetry and the
 Chazy equation, {\em J. Diff. Eq.} {\bf 124}  (1994), 225--246.
\bibitem{olver} Gonz\'alez-L\'opez A, Kamran N and  Olver PJ,
 Lie algebras of vectors fields in the real plane,
 {\em Proc. London Math. Soc.} {\bf 64} (1992), 339--368.
 \bibitem{Hawkins} Hawkins T, Jacobi and the birth of Lie's theory
 of groups, {\em  Arch. Hist. Exact Sciences} {\bf 42} (1991),
 187--278.
 \bibitem{nailio} Ibragimov NH and  Nucci MC, Integration of third order ordinary differential equations by Lie's method: equations admitting
 three-dimensional Lie algebras, {\em Lie Groups and their Applications} {\bf 2} (1994), 49--64.
 \bibitem{Lie5}
Lie S,  Vorlesungen \"uber Differentialgleichungen mit Bekannten
Infinitesimalen Transformationen, B G Teubner, Leipzig,  1912.
\bibitem{mahomed}  Mahomed FM, Symmetry Lie algebras of
 n$^{th}$ order ordinary differential equations,
 PhD Thesis, University of the Witwatersrand, Johannesburg, 1989.
\bibitem{Reduce} Nucci MC, Interactive REDUCE programs
 for calculating Lie point, non-classical, Lie-B\"acklund, and approximate symmetries of
 differential equations: manual and floppy disk, in  CRC Handbook of Lie Group Analysis of
Differential Equations, Vol. 3: New Trends in Theoretical
Developments and Computational Methods, Editor: Ibragimov NH, CRC
Press, Boca Raton, 1996, 415--481.
\bibitem{PW} Patera J and Winternitz P, Subalgebras of real
 three- and four-dimensional Lie algebras, {\em J. Math. Phys.} {\bf 18} (1977), 1449--1455.


\end{thebibliography}
\end{document}